\documentclass[aps,pra,10pt,showpacs,amsmath,amssymb,floatfix,superscriptaddress,longbibliography,twocolumn,eqsecnum]{revtex4-2}
\usepackage{physics}
\usepackage{graphicx}
\usepackage[usenames,dvipsnames,svgnames,table]{xcolor}
\usepackage{tcolorbox}
\usepackage{amsthm}
\usepackage[unicode=true,
            pdfusetitle,
            bookmarks=true,
            bookmarksnumbered=false,
            bookmarksopen=false,
            breaklinks=true,
            pdfborder={0 0 0},
            backref=false,
            colorlinks=true,
            hypertexnames=false]{hyperref}
\hypersetup{linkcolor=NavyBlue,urlcolor=NavyBlue,citecolor=NavyBlue}
\newcommand{\qfi}{\mathcal{F}}
\newcommand{\qfiFirst}{\mathcal{F}^{(1)}}
\newcommand{\qfiSecond}{\mathcal{F}^{(2)}}
\newcommand{\qfiThird}{\mathcal{F}^{(3)}}
\newtheorem{prop}{Proposition}

\begin{document}
\title{Quantum {Cram\'er-Rao} bound for quantum statistical models with parameter-dependent rank}

\author{Yating Ye}
\affiliation{Department of Physics, Hangzhou Dianzi University, Hangzhou 310018, China}
\author{Xiao-Ming Lu}
\email{lxm@hdu.edu.cn}
\homepage{http://xmlu.me}
\affiliation{Department of Physics, Hangzhou Dianzi University, Hangzhou 310018, China}

\begin{abstract}
Recently, a widely-used computation expression for quantum Fisher information was shown to be discontinuous at the parameter points where the rank of the parametric density operator changes.
The quantum Cram\'er-Rao bound can be violated on such singular parameter points if one uses this computation expression for quantum Fisher information.
We point out that the discontinuity of the computation expression of quantum Fisher information is accompanied with the unboundedness of the symmetric logarithmic derivation operators, based on which the quantum Fisher information is formally defined and the quantum Cram\'er-Rao bound is originally proved.
We argue that the limiting version of quantum Cram\'er-Rao bound still holds when the parametric density operator changes its rank by closing the potential loophole of involving an unbounded SLD operator in the proof of the bound.
Moreover, we analyze a typical example of the quantum statistical models with parameter-dependent rank.
\end{abstract}

\maketitle

\section{Introduction}

The Fisher information matrix characterizes the least mean square error of parameter estimation for a large number of samples~\cite{Fisher1922,Fisher1925}.
For a finite number of samples, the Fisher information matrix also gives a lower bound (namely, the Cram\'er-Rao bound) on estimation errors when the estimator are restricted to be locally unbiased~\cite{Cramer1946,Rao1945}.
These properties make the Fisher information matrix play a pivotal role in classical parameter estimation theory~\cite{Casella2002,Vaart1998,Hayashi2005}.

For quantum parameter estimation problems, not only the classical estimators but also the quantum measurements should be taken into consideration for minimizing the estimation errors.
Helstrom derived the first quantum Cram\'er-Rao bound (QCRB) by defining the quantum Fisher information (QFI) matrix as an analogue of the classical one~\cite{Helstrom1967,Helstrom1968,Helstrom1976,Paris2009,Liu2020}.
Due to measurement incompatibility caused by Heisenberg's uncertainty principle~\cite{Heisenberg1927}, the quantum estimation problems for multiple parameters are much more intricate than the classical estimation problems~\cite{Yuen1973,Holevo2011,Lu2021,Carollo2019a,Belliardo2021,Chen2022,Kull2020,Sidhu2021,Suzuki2019}.
For single-parameter estimation problems, the QCRB, which is given by the QFI, gives a quite satisfactory approach to revealing the ultimate quantum limit of estimation precision.

However, recent researches disclosed a possible defect about the QCRB at the parameter points where the rank of the parametric density operator changes~\cite{Safranek2017,Seveso2019,Zhou2019b,Rezakhani2019,Goldberg2021}.
\v{S}afr\'anek~\cite{Safranek2017} demonstrated that a widely-used computation expression of QFI can be different from the Bures metric, which had been considered to be equivalent (up to a constant factor) to the QFI for all cases, and discontinuous at such singular parameter points.
Later, Seveso \textit{et al}.~\cite{Seveso2019} showed through examples that the QCRB based on the QFI does not hold at these limiting cases. 
However, it is still not clear why the QCRB can be violated and where is the loophole in the previous proof of the QCRB.
This motivate us to study in detail the QCRB in the quantum parameter estimation problems with parameter-dependent rank.

In this work, we revisit the derivation of the QCRB and analyze the potential loophole that may cause the failure of the QCRB.
We point out that the symmetric logarithmic derivative (SLD) operator, through which the QCRB is derived and the QFI is defined, is implicitly assumed to be bounded in the previous derivation of the QCRB as well as in the computation of the QFI.
When the value of parameter approaches the singular parameter points where the rank of density operator changes, the SLD operator may become unbounded and some previously-used computation expressions of the QFI might inconsistent with the definition of the QFI.
We show that this potential loophole can be closed by proving the QCRB in a manner that the SLD operator is not involved.
Moreover, we analyze in detail the QCRB in some typical examples of the quantum statistical models with parameter-dependent rank.

This paper is organized as follows. 
In Sec.~\ref{sec:qfi_review}, we give a brief review on the formal definition of the QFI and two expressions for the QFI. 
In Sec.~\ref{sec:disc_analysis}, we investigate the relationships between the discontinuity of the QFI and the boundedness of the SLD operator.
In Sec.~\ref{sec:new_qcrb}, we discuss on how to prove the QCRB without resorting to the SLD operator.
In Sec.~\ref{sec:example}, we analyze in detail the validity of the QCRB with typical examples.
We summarize our results in Sec.~\ref{sec:conclusion}.

\section{Expressions for the QFI and the discontinuity}
\label{sec:qfi_review}

Let us start by considering the general problem of estimating a single parameter with quantum systems.
Assume that the state of the quantum system, described by a density operator \(\rho_\theta\), depends on an unknown parameter \(\theta\).
The value of \(\theta\) can be estimated by data-processing on outcomes obtained from a measurement performed on the quantum system.
A quantum measurement is mathematically characterized by a positive-operator-valued measure (POVM) \(\{E_x \mid E_x \geq 0, \sum_x E_x = \openone\}\) with \(x\) denoting the outcome and \(\openone\) being the identity operator.
According to quantum mechanics, the probability of obtaining an outcome \(x\) is given by \(\tr(\rho_\theta E_x)\).
The data-processing is represented by an estimator \(\hat\theta\) that maps the measurement outcome \(x\) into the estimates for \(\theta\).
The estimation error can be assessed by the mean square error defined by
\begin{equation}
	\mathcal E_\theta :=  \mathbb E_\theta[(\hat\theta -\theta)^2]
	= \sum_x [\hat \theta(x) - \theta ]^2 \tr(\rho_\theta E_x),
\end{equation}
where \(\mathbb E_\theta\) denotes the expectation under the probability distribution given by \(p_\theta(x) = \tr(\rho_\theta E_x)\).
For all quantum measurements and all unbiased estimators \(\hat\theta\) that satisfy
\begin{align}
	\mathbb E_\theta[\hat\theta] = \sum_x \hat\theta(x) \tr(\rho_\theta E_x) = \theta
\end{align}
for all possible true values of \(\theta\), the estimation error obeys the QCRB (also known as the Helstrom bound):
\begin{equation}\label{eq:qcrb}
	\mathcal E_\theta \geq \frac1{\qfi_\theta},
\end{equation}
where \(\qfi_\theta\) is the QFI whose definition will be given in what follows.

There exist three expressions for the QFI that are often used in the previous works~\cite{Helstrom1967,Helstrom1968,Paris2009,Braunstein1994}.
The first one is its formal definition~\cite{Helstrom1967,Helstrom1968}:
\begin{equation}\label{eq:F1}
	\qfiFirst_\theta := \tr(\rho_\theta L_\theta^2),
\end{equation}
where \(L_\theta\) is the SLD operator defined as the Hermitian operator satisfying 
\begin{equation} \label{eq:sld}
	\frac{1}{2} (L_\theta \rho_\theta +\rho_\theta L_\theta)  = \dv{\rho_\theta }{\theta}.	
\end{equation}
The QCRB \(\mathcal E_\theta \geq 1 / \qfiFirst_\theta\) was originally proved by utilizing Eq.~\eqref{eq:F1} and Eq.~\eqref{eq:sld}.

Suppose that \(\rho_\theta=\sum_j \lambda_j \op{e_j}\) is the spectral decomposition of \(\rho_\theta\).
Note that the eigenvalues \(\lambda_j\) and the eigenvectors \(\ket{e_j}\) depends on the true value of the parameter \(\theta\).
The second form of the QFI is a widely-used computation expression given by
\begin{align} \label{eq:F2}
	\qfiSecond_\theta &:= \sum_{j,k \in K_\theta} 
	\frac{2}{\lambda_j + \lambda_k} 
	\abs{\mel{e_j}{\dv{\rho_\theta }{\theta}}{e_k}}^2,
\end{align}
where \(K_{\theta}\) is the index set \(\{k \mid \lambda_k(\theta) = 0\}\) with \(\lambda_j(\theta)\) denoting the value of \(\lambda_j\) at the parameter point \(\theta\).
This expression can be derived by solving the matrix representation of the SLD operator \(L_\theta\) with the basis constituted by the eigenvectors of \(\rho_\theta\) and then substituting it into the formal definition Eq.~\eqref{eq:F1} of the QFI.

The third form of the QFI is given by
\begin{equation} \label{eq:F3}
	\qfiThird_\theta
	:= 8 \lim_{\epsilon\to0}\frac{1 - \norm{\sqrt{\rho_\theta}\sqrt{\rho_{\theta+\epsilon}}}_1}{\epsilon^2},
\end{equation}
where \( \norm{X}_p :=\qty[\tr(|X|^p)]^{1/p}\) with \( |X|:=\sqrt{X^{\dagger}X} \) is the Schatten-\(p\) norm of an operator \(X\). 
The quantity \(\norm{\sqrt{\rho_\theta}\sqrt{\rho_{\theta+\epsilon}}}_1\) is known as Uhlmman's fidelity between the density operators \(\rho_\theta\) and \(\rho_{\theta+\epsilon}\)~\cite{Uhlmann1976,Uhlmann2011,Jozsa1994}.
This expression is deeply relevant to the Bures distance~\cite{Bures1969} between two density operators $\rho$ and $\sigma$: 
\begin{equation}
	d_{\mathrm{B}}(\rho,\sigma)
	:= \sqrt{
		2 \qty( 1 - \norm{\sqrt\rho \sqrt\sigma}_1)
	}.
\end{equation}
It can be seen that the third form \(\qfiThird_\theta\) is equivalent to the metric of the Bures distance between two density operators up to an insignificant constant factor \(4\), i.e.,
\begin{equation}
	\qfiThird_\theta
	= 4 \lim_{\epsilon\to0} 
		\frac{d_{\mathrm{B}}(\rho_\theta,\rho_{\theta + \epsilon})^2}{\epsilon^2}.
\end{equation}
In Ref.~\cite{Zhou2019b}, Zhou and Jiang showed that the computation expression \(\qfiSecond_\theta\) can be fully equivalent to a modified Bures metric: 
\begin{equation}
	\qfiSecond_\theta = 4 \lim_{\epsilon\to0} 
		\frac{d_{\mathrm{B}}(\rho_{\theta - \epsilon /2},\rho_{\theta + \epsilon/2})^2}{\epsilon^2}.
\end{equation}

\section{Discontinuous QFI and unbounded SLD}
\label{sec:disc_analysis}
As shown by \v{S}afr\'anek~\cite{Safranek2017}, the expressions \(\qfiSecond_\theta\) and \(\qfiThird_\theta\), given in Eqs.~\eqref{eq:F2} and \eqref{eq:F3}, can be different at the parameter points where the rank of \(\rho_\theta\) changes.
The discrepancy occurs when a term excluded from the summation in Eq.~\eqref{eq:F3} for \(\qfiSecond_\theta\) has a finite value as the parameter tends to that specific value. 
At a specific value \(\theta'\), it has been shown that~\cite{Safranek2017}
\begin{align} \label{eq:Delta}
	\Delta_{\theta'} 
	& := \qfiThird_{\theta'} - \qfiSecond_{\theta'}
	   = \lim_{\theta \to \theta'} \sum_{k \in K_{\theta'}} 
		 \frac{1}{\lambda_k} \qty(\dv{\lambda_k}{\theta})^2 \nonumber \\
	&  = \sum_{k \in K_{\theta'}} 
		 2 \evaluated{\dv[2]{\lambda_k}{\theta}}_{\theta=\theta'},
\end{align}
where \(\lambda_j\) are the eigenvalues of \(\rho_\theta\) and \(K_{\theta'}\) denotes the index set \(\{k \mid \lambda_k(\theta') = 0\}\).
Note that the last equality in Eq.~\eqref{eq:Delta} is due to L'H\^opital's rule in calculus~\cite{Safranek2017,Seveso2019}.
The \(\qfiSecond\)-QCRB can be violated when \(\Delta_{\theta} > 0\), see Ref.~\cite{Seveso2019}. 
In what follows, we shall interpret why the \(\qfiSecond\)-QCRB breaks down at the parameter point where the rank of the density operators changes.

First, when the SLD operator is bounded, the \(\qfiSecond\)-QCRB always holds.
This is because the original proof of \(\qfiFirst\)-QCRB and the relation \(\qfiSecond_\theta=\qfiFirst_\theta\) are both rigorous for such case.
To see the latter, recall that the expression Eq.~\eqref{eq:F2} for \(\qfiSecond_\theta\) is a consequence of solving the SLD equation Eq.~\eqref{eq:sld} for \(L_\theta\) through matrix representation with the basis constituted by the eigenvectors of \(\rho_\theta\).
With the spectral decomposition \(\rho_\theta=\sum_j \lambda_j \op{e_j}\), it follows from Eq.~\eqref{eq:sld} that
\begin{equation} \label{eq:sld_element}
	\frac{\lambda_j + \lambda_k}{2} \mel{e_j}{L_\theta}{e_k} 
	= \mel{e_j}{\dv{\rho_\theta }{\theta}}{e_k}.
\end{equation}
When \(\lambda_j + \lambda_k \neq 0\), the above equality implies that
\begin{equation} \label{eq:L-jk}
	 \mel{e_j}{L_\theta}{e_k} 
	= \frac{2}{\lambda_j + \lambda_k} \mel{e_j}{\dv{\rho_\theta }{\theta}}{e_k}.
\end{equation}
For those indices \(j\) and \(k\) such that \(\lambda_j + \lambda_k = 0\), \(\mel{e_j}{L_\theta}{e_k} \) is indefinite by Eq.~\eqref{eq:sld_element}.
On the other hand, the QFI defined in Eq.~\eqref{eq:F1} can be expressed with the spectral decomposition of \(\rho\) as
\begin{align} 
	\qfiFirst_\theta &= \sum_{jk} \lambda_j \abs{\ev{e_j|L_\theta|e_k}}^2 \nonumber \\
	&= \sum_{jk} \frac{\lambda_j + \lambda_k}{2} \abs{\ev{e_j|L_\theta|e_k}}^2, \label{eq:qfi_subs}
\end{align}
where the second equality is due to the symmetrization with respect to the indices \(j\) and \(k\).
For those indices \(j\) and \(k\) such that \(\lambda_j + \lambda_k = 0\), the corresponding terms in the right hand side of Eq.~\eqref{eq:qfi_subs} have no contribution to the summation as long as the SLD operator is \emph{bounded}.
In such case, combining Eqs.~\eqref{eq:sld_element} and \eqref{eq:qfi_subs}, we can see that \(\qfiFirst_\theta = \qfiSecond_\theta\).
Therefore, the QCRB based on \(\qfiSecond_\theta\) must hold when there exists an bounded SLD operator.

It is not always possible to have a bounded SLD operator~\cite{Uhlmann1991} for all range of the parameter, even when the QFI is finite.
For example, for
\begin{equation}\label{eq:example}
	\rho_\theta = \cos^2\theta \op0 + \sin^2\theta \op1 
	\mbox{ with }
	0 \leq \theta \leq \pi / 2,
\end{equation}
the SLD operator is 
\begin{equation}
	L_\theta = -2 \tan\theta\op0 + 2 \cot\theta\op1,	
\end{equation}
which is unbounded at \(\theta=0\) and \(\theta=\pi/2\).
Meanwhile, the QFI for the above statistical model is \(4\) for all \(0 \leq \theta \leq \pi / 2\).

As shown by Ref.~\cite{Safranek2017}, \(\qfiThird_\theta\) represents the continuous version of the QFI \(\qfiFirst_\theta\), i.e., \(\lim_{\theta\to\theta'} \qfiFirst_\theta=\qfiThird_{\theta'}\).
Therefore, we can substitute Eq.~\eqref{eq:qfi_subs} into the definition of \(\Delta_{\theta'}\) and thus get
\begin{equation}
	\Delta_{\theta'} = \lim_{\theta \to \theta'} \sum_{j,k \in K_{\theta'}} 
	\frac{\lambda_j + \lambda_k}{2} \abs{\ev{e_j|L_\theta|e_k}}^2.
\end{equation}
Since \((\lambda_j + \lambda_k) / 2 = 0\) at the parameter point \(\theta'\) for all \(j,k \in K_{\theta'} \), we get
\begin{equation}
	\Delta_{\theta'} > 0 
	\iff
	\lim_{\theta\to\theta'} L_\theta \mbox{ is unbounded},
\end{equation}
meaning that the discrepancy between \(\qfiSecond_\theta\) and \(\qfiThird_\theta\) must be accompanied by an unbounded SLD operator.
The divergence of the SLD operator has also been investigated through the integral representation of the SLD operator by Rezakhani, Hassani, and Alipour in Ref.~\cite{Rezakhani2019}.

Strictly speaking, it is not the QFI (defined by \(\qfiFirst_\theta\)) but its computation expression \(\qfiSecond_\theta\) that is discontinuous when the density operator changes its rank. 
At these singular parameter points, \(\qfiSecond_\theta\) does not coincide with the definition of the QFI and thus the \(\qfiSecond\)-QCRB, i.e., \(\mathcal E_\theta \geq 1 / \qfiSecond_\theta\), can be violated.

\section{QCRB at rank-changing points}
\label{sec:new_qcrb}
Even though the violation of the \(\qfiSecond_\theta\) can be ascribed to the inconsistency between \(\qfiSecond_\theta\) and \(\qfiFirst_\theta\) when the SLD operator is unbounded, it does not mean that the limiting version of \(\qfiFirst\)-QCRB, namely,
\begin{equation} \label{eq:continuous_qcrb}
	\mathcal E_{\theta'} \geq \frac1{\lim_{\theta \to \theta'}\qfiFirst_\theta},
\end{equation}
has been rigorously proved at such singular parameter point \(\theta'\).
This is because the original proof of QCRB~\cite{Helstrom1967,Helstrom1968} did not discuss the cases of unbounded SLD operators.
We have to be cautious about whether some mathematical formulas used in the original proof of the QCRB still hold for unbounded operators, e.g., the cyclic property of the trace operation, \(\tr(AB)=\tr(BA)\) for two operators \(A\) and \(B\), can be violated if unbounded operators are involved.  
We here argue that the limiting version of \(\qfiFirst\)-QCRB holds even when the SLD operator becomes unbounded and elucidate it from two perspectives in what follows.

Our first approach to supporting Eq.~\eqref{eq:continuous_qcrb} is adapting the original proof of the QCRB~\cite{Helstrom1967,Helstrom1968,Helstrom1974} to avoid involving the unbounded SLD.
Note that the influence of the SLD operator is always imposed via the product \(L_\theta \sqrt{\rho_\theta}\).
When the SLD operator is unbounded, the product \(L_\theta \sqrt{\rho_\theta}\) may still be bounded.
For example, for the density operators in Eq.~\eqref{eq:example}, even the SLD operators is unbounded at \(\theta=0\) and \(\theta=\pi/2\), the product
\begin{equation}
	L_\theta\sqrt{\rho_\theta} = -2 \sin\theta\op0 + 2 \cos\theta\op1
\end{equation}
is still bounded at \(\theta=0\) and \(\theta=\pi/2\).
This motives us to introduce the linear operator \(Q_\theta\) that satisfies the following two conditions:
\begin{align}
	\frac12\qty(Q_\theta \sqrt{\rho_\theta} + \sqrt{\rho_\theta} Q_\theta^\dagger) 
	&= \dv{\rho_\theta }{\theta}, \label{eq:Q1}\\
	\sqrt{\rho_\theta} Q_\theta - Q_\theta^\dagger \sqrt{\rho_\theta} &= 0. \label{eq:Q2}
\end{align}
In fact, \(Q_\theta\) plays the role of \(L_\theta \sqrt{\rho_\theta}\), as Eq.~\eqref{eq:Q1} corresponds to the SLD equation Eq.~\eqref{eq:sld} and Eq.~\eqref{eq:Q2} corresponds to the hermiticity of the SLD operator.
Moreover, the QFI can always be expressed as
\begin{equation}  \label{eq:QFI_Q}
	\qfiFirst_{\theta} =  \norm{Q_{\theta}}_2^2.
\end{equation}
The advantage of \(Q_\theta\) over the SLD operator \(L_\theta\) is that, according to Eq.~\eqref{eq:QFI_Q}, the QFI is infinitely large if and only if \(Q_\theta\) is unbounded.
Therefore, if \(Q_\theta\) is unbounded, the \(\qfiFirst\)-QCRB becomes \(\mathcal E_\theta \geq 0\), whose validation is trivial.
So we only need to consider the cases where \(Q_\theta\) are bounded.

Now, we assume that \(Q_\theta\) is bounded henceforth and prove the QCRB without directly resorting to the SLD operator.
For any POVM \(\qty{E_x}\) and any estimator \(\hat\theta\),  let us define \(X := \sum_x \hat \theta(x) E_x\) and \(\Delta X := X - \tr(\rho_\theta X) \openone\), where \(\openone\) denotes the identity operator.
The mean of the estimator is given by \(\mathbb E_\theta[\hat \theta] = \tr(\rho_\theta X)\).
The variance of the estimator, \(\mathcal V_\theta := \mathbb E[\hat\theta^2] - \mathbb E[\hat\theta]^2\), is bounded from below by the variance of the operator \(X\)~\cite[see Section 2.9]{Holevo2011}, that is,
\begin{equation} \label{eq:proof1}
	\mathcal V_\theta \geq \tr[\rho (\Delta X)^2] = \norm{(\Delta X)\sqrt{\rho_\theta}}_2^2.
\end{equation}
Applying the Cauchy-Schwarz inequality, it follows that
\begin{align} \label{eq:proof2}
	\norm{(\Delta X) \sqrt{\rho_\theta}}_2 \norm{Q_\theta}_2
	\geq |\tr[\sqrt{\rho_\theta}(\Delta X) Q_\theta]|.
\end{align}
Since \(|z|\geq |\Re z|\) for any complex number \(z\), we get
\begin{align}
	|\tr[\sqrt{\rho_\theta}(\Delta X) Q_\theta]| 
	\geq\ & |\Re\tr[\sqrt{\rho_\theta}(\Delta X) Q_\theta]| \nonumber\\
	=\ & \qty|\tr(\Delta X \frac{Q_\theta \sqrt{\rho_\theta} + \sqrt{\rho_\theta} Q_\theta^\dagger}{2})|  \nonumber \\
	=\ & \qty|\tr(\Delta X \dv{\rho_\theta}{\theta})| \nonumber \\
	=\ & \qty|\dv{\tr(\rho_\theta X)}{\theta}| = \qty|\dv{\mathbb E_\theta[\hat\theta]}{\theta}|,  
	\label{eq:proof3}
\end{align}
where the equality in the second last line is due to Eq.~\eqref{eq:Q1}.
It then follows from Eqs.~(\ref{eq:proof1}--\ref{eq:proof3}) that 
\begin{equation} \label{eq:qcrb_biased}
	\mathcal V_\theta 
	\geq \qty( \dv{\theta} \mathbb E_\theta[\hat\theta] )^2 \norm{Q_\theta}_2^{-2}
	\geq \qty( \dv{\theta} \mathbb E_\theta[\hat\theta] )^2 / \qfiFirst_\theta.
\end{equation}
For locally unbiased estimators, for which we have \(\mathcal E_\theta = \mathcal V_\theta\) and \(\dv{\theta} \mathbb E[\hat\theta]=1\), the inequality in Eq.~\eqref{eq:qcrb_biased} becomes \(\mathcal E_\theta \geq \norm{Q_\theta}_2^{-2}\).

Using the operator \(Q_\theta\) instead of the SLD operator \(L_\theta\), we have proved that the QCRB \(\mathcal E_\theta \geq \norm{Q_\theta}_2^{-2}\) always holds no matter the operator \(Q_\theta\) is bounded or not.
This proof can include the special case where \(L_\theta\) is unbounded at a singular parameter point \(\theta'\) but the QFI is still finite.
In such cases, the QFI can be expressed as \(\norm{Q_{\theta'}}_2^2 = \lim_{\theta\to\theta'}\tr(\rho_\theta L_\theta^2)\) due to the relation between \(Q_\theta\) and \(L_\theta\).
Therefore, we get Eq.~\eqref{eq:continuous_qcrb}.

Our second approach to supporting the limiting version of the  \(\qfiFirst\)-QCRB is bounding the estimation error with the Bures metric, namely, Eq.~\eqref{eq:qcrb_qfi3}.
It is known that the Bures metric satisfies \(\qfiThird_{\theta'} = \lim_{\theta\to\theta'} \qfiFirst_\theta\), e.g. see Ref.~\cite{Safranek2017}, so the inequality Eq.~\eqref{eq:continuous_qcrb} is equivalent to
\begin{equation} \label{eq:qcrb_qfi3}
 	\mathcal E_{\theta'} \geq \frac1{\qfiThird_{\theta'}}.
\end{equation} 
This inequality was recently proved by Yang, Chiribella, and Hayashi through a more fundamental relation (denoted by Yang-Chiribella-Hayashi inequality henceforth)~\cite{Yang2019a}: 
\begin{equation}\label{eq:YCH}
	\frac12 \qty(\mathcal E_\theta + \mathcal E_{\theta+\epsilon} + \epsilon^2)
	\geq \frac{\epsilon^2}{4 d_\mathrm{B}(\rho_\theta, \rho_{\theta+\epsilon})^2}
\end{equation}
for the estimators that are unbiased at both \(\theta\) and \(\theta+\epsilon\). 
Taking the limit \(\epsilon\to0\) of the above inequality, Eq.~\eqref{eq:qcrb_qfi3} can be obtained.
We also give an alternative proof by using the purification method in Appendix~\ref{app:purification}.
In addition, the inequality Eq.~\eqref{eq:YCH} can be easily generalized to include the biased estimators, as shown in Appendix~\ref{app:YCH}.

\section{Concrete examples}
\label{sec:example}
In Ref.~\cite{Seveso2019}, Seveso \textit{et al}.~argued with examples that not only the  \(\qfiSecond\)-QCRB but also the \(\qfiThird\)-QCRB can be violated when the parametric density operator changes its rank through examples.
This conflicts with our result Eq.~\eqref{eq:continuous_qcrb} and Eq.~\eqref{eq:qcrb_qfi3} in Sec.~\ref{sec:new_qcrb}.
In what follows, we shall analyze in details the example discussed in Ref.~\cite{Seveso2019} and solve the conflict.

Following Ref.~\cite{Seveso2019}, we first consider the example whose parametric density operator reads
\begin{equation}\label{eq:example1}
	\rho_q = (1 - q) \op0 + q \op1 \mbox{ with } 0 \leq q \leq 1
\end{equation}
with \(q\) being the parameter of interest.
The rank of \(\rho_q\) changes at the boundary, i.e., \(q=0\) and \(q=1\).
The SLD is given by 
	\begin{equation}
		L_q = \frac{-1}{1-q}\op0 + \frac{1}{q} \op1,
	\end{equation}
which becomes unbounded as \(q \to 0\) or \(q \to 1\).
Meanwhile, the Bures distance is given by
\begin{equation}
	d_\mathrm{B}(\rho_q,\rho_{q'})^2 
	= 2 \qty[1 - \sqrt{(1-q)(1-q')} - \sqrt{q q'}].
\end{equation}
The QFI expressions are given by~\cite{Seveso2019}
\begin{equation} \label{eq:example1_qfi3}
	\qfiThird_q
	= \frac1{q(1-q)},
\end{equation}
which diverges at \(q=0\) and \(q=1\), and
\begin{equation} \label{eq:example1_qfi2}
	\qfiSecond_q = \begin{cases}
		1, &\text{ if } q=0 \text{ or } 1, \\
		\frac1{q(1-q)}, & 0 < q < 1.
	\end{cases}
\end{equation}
Suppose that we have \(n\) copies of the quantum system and perform on each copy the projective measurement whose POVM is given by \(E_0 = \op0\) and \(E_1=\op1\).
Denoted by \(x_j\) the measurement outcome on \(j\)-th system, which takes the value \(0\) and \(1\) with the probabilities \(1-q\) and \(q\), respectively.
The statistical quantity
\begin{equation}\label{eq:t}
	t := \sum_{j=1}^n x_j
\end{equation}
is a sufficient statistic for estimating \(q\).
The probability of \(t\) is given by
\begin{equation}\label{eq:pq}
	p_q(t) = \binom{n}{t} q^{t} (1-q)^{n - t}.
\end{equation}
Note that
\begin{equation}
	\pdv{\ln p_q(t)}{q} 
	= \frac{t}{q} - \frac{n - t}{1 - q}
	= \frac{t - nq}{q (1 - q)},
\end{equation}
implying that the maximum likelihood estimator for this statistical model is given by 
\begin{equation}
	\hat q_\mathrm{ML}(t)  = \frac{t}{n}.
\end{equation}
The mean of \(\hat q_\mathrm{ML}(t)\) is \(q\), so \(\hat q_\mathrm{ML}(t)\) is unbiased for all \(0\leq q\leq 1\).
The variance of \(\hat q_\mathrm{ML}(t)\) is \(q(1 - q) / n\), so the mean-square error of the maximum likelihood estimator vanishes at \(q=0\) and \(q=1\).
Comparing the meas-square error of the maximum likelihood estimation with the QFI expressions Eq.~\eqref{eq:example1_qfi3} and Eq.~\ref{eq:example1_qfi2}, it can be seen that the \(\qfiSecond\)-QCRB is violated at \(q=0\) and \(q=1\) while the \(\qfiThird\)-QCRB still holds in such a scenario~\cite{Seveso2019}.

Furthermore, Ref.~\cite{Seveso2019} considered a reparametrization of the above example, namely,
\begin{equation} \label{eq:example2}
	\rho_\vartheta = \cos^2\vartheta \op0 + \sin^2\vartheta \op1
	\mbox{ with } 0 \leq \vartheta \leq \frac{\pi}{2}.
\end{equation}
This is a reparametrization of \(\rho_q\) in Eq.~\eqref{eq:example1} by substituting \(q = \sin^2\vartheta\) therein.
For this statistical model, the SLD is given by 
\begin{equation}
	L_\vartheta	= -2 \tan\vartheta \op0 + 2 \cot\vartheta \op1,
\end{equation}
which becomes unbounded at \(\vartheta = 0\) and \(\vartheta = \pi / 2\).
Correspondingly, the QFI are given by~\cite{Safranek2017}
\begin{equation}
	\qfiThird_\vartheta = 4 \qand
	\qfiSecond_\vartheta = \begin{cases}
		0, &\text{ if } \vartheta=0 \text{ or } \pi / 2, \\
		4, &\text{ otherwise. }
	\end{cases}
\end{equation}
The optimal measurement for estimating \(\vartheta\) is still given by the POVM \(\qty{\op0,\op1}\), as the parametric density matrix is always diagonal with the basis \(\qty{\ket0, \ket1}\).
Because the maximum likelihood estimator is equivariant~\cite[Section 9.4]{Wasserman2010}, the maximum likelihood estimator \(\hat\vartheta_\mathrm{ML}\) for \(\vartheta\) can be expressed as 
\begin{equation} \label{eq:MLE_equivariant}
	\hat\vartheta_\mathrm{ML}(t) 
	= \arcsin(\sqrt{\hat q_\mathrm{ML}(t) })
	= \arcsin(\sqrt\frac{t}{n}),
\end{equation}
where \(t\) is the same as that in Eq.~\eqref{eq:t}.
This is equivalent to first estimate \(q\) in Eq.~\eqref{eq:example1} using the maximum likelihood estimation and then calculate \(\vartheta\) via the relation \(q=\sin^2\vartheta\). 
It can be seen from Eq.~\eqref{eq:pq} that \(t\) has a deterministic value at \(q=0\) and \(q=1\), which correspond to \(\vartheta=0\) and \(\vartheta = \pi/2\), respectively.
Consequently, the variance of \(\hat\vartheta_\mathrm{ML}\) vanishes at these parameter points.
Meanwhile, the continuous version of QCRB reads \(\mathcal E_\vartheta \geq 1/(4n)\).
Due to these facts, it was concluded in Ref.~\cite{Seveso2019} that the continuous version of QCRB (or the \(\qfiThird\)-QCRB) is violated at \(\theta=0\) and \(\theta = \pi / 2\).
However, the continuous version of QCRB should be valid according to our analysis in the previous section.
We shall solve this conflict in what follows.

\begin{figure}[tb]
	\centering
	\includegraphics[width=\linewidth]{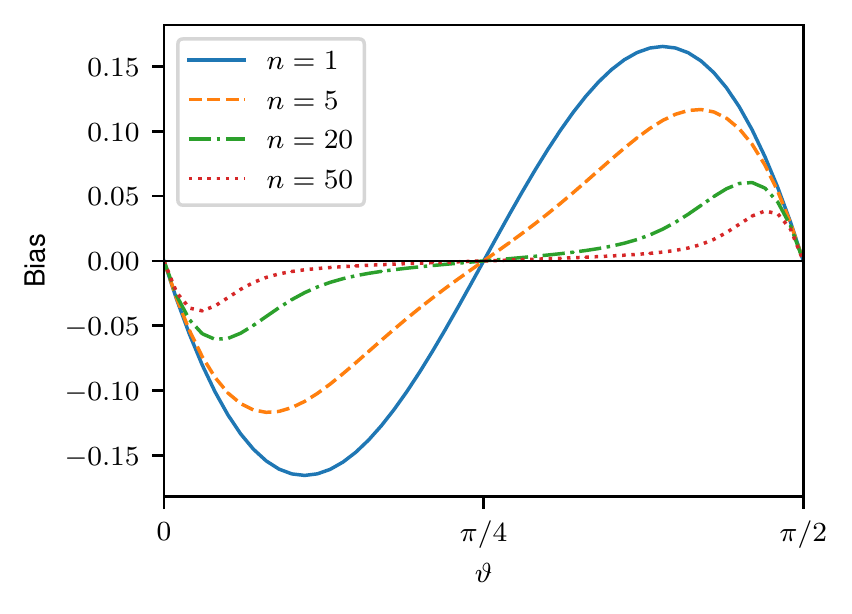}
	\caption{
		The bias, \(\mathbb E_\vartheta[\hat\vartheta_\mathrm{ML}] - \vartheta\), of the maximum likelihood estimator \(\hat\vartheta_\mathrm{ML}\) versus the true value \(\vartheta\).
		Here, \(n\) is the number of samples used for estimation.
	}
	\label{fig:bias}
\end{figure}

Note that the QCRB is established only for \emph{locally unbiased estimators}.
Although \(\hat q_\mathrm{ML}\) is unbiased for \(q\), \(\hat \vartheta_\mathrm{ML}\), as we will show in the following, is not locally unbiased for \(\vartheta\).
The mean of the maximum likelihood estimator is given by
\begin{equation}
	\mathbb E_\vartheta[\hat\vartheta_\mathrm{ML}(t)] 
	= \sum_{t=0}^n \arcsin(\sqrt\frac{t}{n}) p_q(t)
\end{equation}
with \(q=\sin^2\vartheta\) and \(p_q(t)\) being given by Eq.~\eqref{eq:pq}.
Taking the derivative of \(\mathbb E_\vartheta[\hat\vartheta_\mathrm{ML}(t)]\), we have
\begin{align}
	& \dv{\vartheta} \mathbb E_\vartheta[\hat\vartheta_\mathrm{ML}(t)] 
	= \sum_{t=0}^n \arcsin(\sqrt\frac{t}{n}) \dv{q}{\vartheta} \eval{\pdv{p_q(t)}{q}}_{q=\sin^2\vartheta} \nonumber \\
	&= 2 \sin\vartheta\cos\vartheta \sum_{t=0}^n \arcsin(\sqrt\frac{t}{n}) \eval{\pdv{p_q(t)}{q}}_{q=\sin^2\vartheta}.
\end{align}
At \(\vartheta = 0\) and \(\vartheta=\pi/2\), it can be seen that \(\dv{\vartheta} \mathbb E_\vartheta[\hat\vartheta_\mathrm{ML}(t)]\) vanishes  for a finite \(n\), as \(\sin\vartheta\cos\vartheta=0\).
Nevertheless, the locally unbiased condition requires that \(\dv{\vartheta} \mathbb E_\vartheta[\hat\vartheta_\mathrm{ML}(t)] = 1\).
Therefore, \(\hat\vartheta\) is not locally unbiased at \(\vartheta = 0\) and \(\vartheta=\pi/2\).
Figure~\ref{fig:bias} plots the bias of the maximum likelihood estimator with \(n\) samples.
It can be seen that the bias abruptly increases when the true value of \(\vartheta\) departs from \(0\) or \(\pi/2\).
Now, it is clear that the fact \(\hat\vartheta_\mathrm{ML}\) has zero variance does not mean the QCRB must be violated, as \(\hat\vartheta_\mathrm{ML}\) is not a locally unbiased estimator.

\begin{figure}[tb]
	\centering
	\includegraphics[width=1\linewidth]{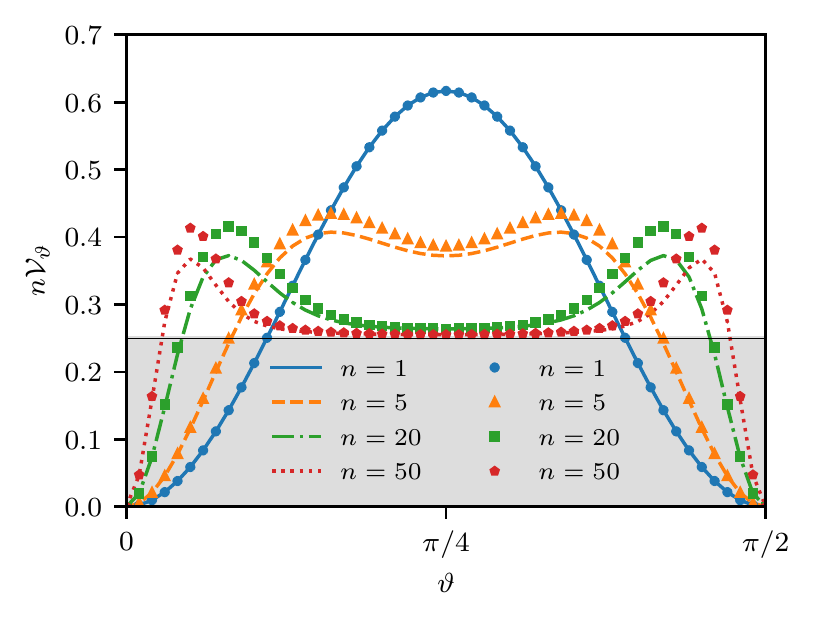}
	\caption{Rescaled variances \(n\mathcal V_\vartheta\) (indicated with various markers) of the maximum likelihood estimator and corresponding biased QCRBs (indicated with various lines).
	Here, \(n\) is the number of samples used for estimation.
	The shadow represent the region where the QCRB for locally unbiased estimator is violated.
	}
	\label{fig:variance}
\end{figure}

Furthermore, we use the biased version of QCRB Eq.~\eqref{eq:qcrb_biased} to verify the continuous QCRB for \(\rho_\vartheta\) at the singular parameter points \(\vartheta=0\) and \(\vartheta=\pi/2\). 
Figure \ref{fig:variance} plots the numerical result of the rescaled variance of the maximum likelihood estimator (\(n \mathcal V_\vartheta\)) and the biased versions of QCRB with different number \(n\) of samples at different true values of \(\vartheta\).
Note that, unlike the QCRB for locally unbiased estimator, the biased version of QCRB depends on the estimators themselves. 
It can be seen from Fig.~\ref{fig:variance} that the biased QCRB always holds, while the unbiased QCRB is violated near the singular parameter points \(\vartheta=0\) and \(\vartheta = \pi / 2\).

\section{Conclusion}
\label{sec:conclusion}
In this work, we have studied the validity of the QCRB at the specific parameter values where the parametric density operator changes its rank.
The violation of the QCRB based on the widely-used computation expression of the QFI, which is  discontinuous at the singular parameter points, is ascribed to the inconsistency between the computation expression of the QFI and the formal definition of the QFI as the SLD operator becomes unbounded.
We have showed that the potential loophole of involving an unbounded SLD operator in the proof of the QCRB can be closed by either using the \(Q_\theta\) operator instead of the SLD operator or invoking the QCRB based on the Bures distance, e.g. the Yang-Chiribella-Hayashi inequality~\cite{Yang2019a}.
We have analyzed in detail the QCRB in a typical example of the quantum statistical models with parameter-dependent rank and solved the conflict with previous results by showing that the maximum likelihood estimator therein is not locally unbiased.
Our work consolidates the soundness of the QCRB as well as the QFI at the rank-changing points.

\begin{acknowledgments}
This work is supported by the National Natural Science Foundation of China (Grants No. 61871162, No. 11935012, and No. 11805048).
\end{acknowledgments}

\appendix
\section{Proof of the QCRB based on the Bures distance by purification}
\label{app:purification}
Although the \(\qfiThird\)-QCRB was already proved by Ref.~\cite{Yang2019a}, we here give an alternative way by utilizing the purification of mixed states.
Let us start by the following proposition.

\begin{prop}\label{prop:purification}
Assume that \(A_\theta\) and \(A_{\theta'}\) are two bounded operators satisfying \(\rho_\theta = A_\theta A_\theta^\dagger\) and \(\rho_{\theta'} = A_{\theta'} A_{\theta'}^\dagger\), respectively.
For any POVM \(\qty{E_x}\) and any unbiased estimator \(\hat\theta\), we have
	\begin{equation}\label{eq:lower_bounds}
		\mathcal E_\theta \geq \frac14 \beta^2 \norm{A_{\theta'} - A_\theta}_2^{-2},
	\end{equation}
	where \(\norm{X}_2 = \sqrt{\tr(X^\dagger X)}\) is the Schatten-\(2\) norm of an operator \(X\) and \(\beta\) is a scalar quantity defined by
	\begin{equation}\label{eq:beta}
	\beta: = \theta' - \theta - \sum_x 
		\eta_{x,\theta} \norm{\sqrt{E_x} (A_{\theta'} - A_\theta)}_2^2
	\end{equation}
	with \(\eta_{x,\theta} := \hat\theta(x) - \theta\) for brevity.
\end{prop}

\begin{proof}
The mean square error of estimating \(\theta\) can be expressed as
\begin{equation} \label{eq:cs1}
	\mathcal E_\theta 
	= \sum_x \eta_{x,\theta}^2
	\norm{\sqrt{E_x} A_\theta}_2^2.
\end{equation}
By inserting \(\sum_x E_x = \openone\) into \(\norm{A_{\theta'} - A_\theta}_2^2 \), we get
\begin{align} \label{eq:cs2}
	\norm{A_{\theta'} - A_\theta}_2^2 
	& = \sum_x \norm{\sqrt{E_x} (A_{\theta'} - A_\theta)}_2^2.
\end{align}
It then can be shown that
\begin{align}
	& \mathcal E_\theta^{1/2} \norm{A_{\theta'} - A_\theta}_2 \nonumber\\
	\geq & 
		\sum_x \abs{\eta_{x,\theta}} 
		\norm{\sqrt{E_x} A_\theta}_2 
		\norm{\sqrt{E_x} (A_{\theta'} - A_\theta)}_2
	\nonumber \\
	\geq & \sum_x 
		\abs{\eta_{x,\theta}} \cdot
		\abs{\tr[A_\theta^\dagger E_x (A_{\theta'} - A_\theta)]}
	=: \alpha.
	\label{eq:alpha}
\end{align}
Here, the first inequality is due to a Cauchy-Schwarz inequality 
\( 
	[\sum_x f(x)^2] [\sum_x g(x)^2] \geq [\sum_x f(x) g(x)]^2
\) 
for two real functions \(f(x) = \eta_{x,\theta} \norm{\sqrt{E_x} A_\theta}_2\) and \(g(x) = \norm{\sqrt{E_x} (A_{\theta'} - A_\theta)}_2\), and the second one is due to a Cauchy-Schwarz inequality \(\norm{X}_2 \norm{Y}_2 \geq \abs{\tr(X^\dagger Y)}\) for two bounded operators \(X = \sqrt{E_x} A_\theta\) and \(Y = \sqrt{E_x} (A_{\theta'} - A_\theta)\).
Noting that \(|z|\geq |\Re z|\) for any complex number \(z\) and \(\sum_x |f_x| \geq \qty|\sum_x f_x|\) for any series of scalars \(f_x\), we get
\begin{align*}
	\alpha &\geq 
	\abs{ 
		\sum_x \eta_{x,\theta}\Re\tr[A_\theta^\dagger E_x (A_{\theta'} - A_\theta)]
	} \nonumber \\
	&= 
	\abs{
		\sum_x 
		\frac{\eta_{x,\theta}}{2} 
		\tr[E_x(
			A_{\theta'} A_\theta^\dagger + A_\theta A_{\theta'}^\dagger - 2 A_\theta A_\theta^\dagger 
		)]
	}. 
\end{align*}
Substituting 
\begin{align}
	& \quad A_{\theta'} A_\theta^\dagger + A_\theta A_{\theta'}^\dagger - 2 A_\theta A_\theta^\dagger 
	 \nonumber\\
	& = - (A_{\theta'} - A_\theta)(A_{\theta'} - A_\theta)^\dagger + A_{\theta'} A_{\theta'}^\dagger - A_\theta A_\theta^\dagger
\end{align}
into the right hand side of the above inequality, we get 
\begin{align}
	\alpha &\geq \frac12
	\Big|
		- \sum_x \eta_{x,\theta} \norm{\sqrt{E_x}(A_{\theta'} - A_\theta)}_2^2  \nonumber \\
		& + \sum_x \eta_{x,\theta} \norm{\sqrt{E_x} A_{\theta'}}_2^2
		- \sum_x \eta_{x,\theta} \norm{\sqrt{E_x} A_{\theta}}_2^2
	\Big|. \label{eq:alpha1}
\end{align}
For all unbiased estimators \(\hat\theta\), \(\sum_x \hat\theta(x)\tr(E_x A_\theta A_\theta^\dagger) = \theta\), which implies that
\begin{align} \label{eq:alpha2}
	\sum_x \eta_{x,\theta} \norm{\sqrt{E_x} A_{\theta'}}_2^2
	= \theta' - \theta
\end{align}
for all \(\theta\) and \(\theta'\).
Specifically for \(\theta'=\theta\), we have 
\begin{equation}
	\sum_x \eta_{x,\theta} \norm{\sqrt{E_x} A_{\theta}}_2^2 = 0. \label{eq:alpha3}
\end{equation}
Substituting Eq.~\eqref{eq:alpha2} and Eq.~\eqref{eq:alpha3} into the right hand side of Eq.~\eqref{eq:alpha1}, we get \(\alpha^2 \geq \beta^2/4\).
Therefore, together with Eq.~\eqref{eq:alpha}, we get Eq.~\eqref{eq:lower_bounds}.
\end{proof}

The inequality Eq.~\eqref{eq:lower_bounds} can be further refined from two perspectives.
Firstly, the operator \(A\) satisfying \(\rho=AA^\dagger\) for a given \(\rho\) is not unique. 
All the operators \(A = \sqrt\rho U\) with \(U\) being an arbitrary unitary operator satisfy \(\rho=AA^\dagger\) and are called amplitudes of \(\rho\).
The inequality Eq.~\eqref{eq:lower_bounds} holds for all amplitudes of \(\rho_\theta\) and \(\rho_{\theta'}\);
This supplies us a freedom of choosing specific \(A_\theta\) and \(A_{\theta'}\) to formulate useful lower bounds based on Eq.~\eqref{eq:lower_bounds}.
The Bures-Uhlmman geometry for density operators tell us that~\cite{Uhlmann1991,Bengtsson2006}
\begin{equation}
	\min_{A_\theta,A_{\theta'}} \norm{A_{\theta'} - A_\theta}_2^2 = d_\mathrm{B}(\rho_\theta,\rho_{\theta'})^2
\end{equation}
and the minimum is attained by the amplitudes satisfying 
\begin{equation}
	A_\theta^\dagger A_{\theta'} = A_{\theta'}^\dagger A_\theta \geq 0.
\end{equation}
Two amplitudes satisfying the above condition are called \emph{parallel}~\cite{Uhlmann2009}.
For parallel amplitudes \(A_\theta\) and \(A_{\theta'}\) we have
\begin{equation}
	\mathcal E_\theta \geq \frac{\beta^2}{4 d_\mathrm{B}(\rho_\theta,\rho_{\theta'})^2}.
\end{equation}
Note that the quantity \(\beta\) defined by Eq.~\eqref{eq:beta} still depends on the POVM and estimator. 

Secondly, the inequality Eq.~\eqref{eq:lower_bounds} holds for any two parameter points \(\theta\) and \(\theta'\) in the range of the parameter.
This supplies us another freedom of formulate useful lower bounds on the estimation error at \(\theta\).
Let us consider two infinitesimally neighboring parameter points by defining \(\epsilon := \theta' - \theta \) and taking the limit \(\epsilon \to 0\).
Assuming that \(A_\theta\) is differentiable so that \(A_{\theta'} \approx A_\theta + \epsilon \dv*{A_\theta}{\theta}\).
Therefore, we obtain
\begin{equation}
	\beta \approx \epsilon - \epsilon^2\sum_x \eta_{x,\theta}
		\tr(\dv{A_\theta^\dagger}{\theta} E_x \dv{A_\theta}{\theta})
\end{equation}
and thus Eq.~\eqref{eq:lower_bounds} becomes
\begin{align}
	\mathcal E_\theta 
	&\geq \frac14 \lim_{\epsilon \to 0} \beta^2 \norm{A_{\theta'} - A_\theta}_2^{-2} 
	= \frac14 \norm{\dv{A_\theta}{\theta}}_2^{-2}.
\end{align}
Note that in Proposition~1 is established the estimator that is unbiased at both the parameter point \(\theta\) and \(\theta+\epsilon\);
When taking the limit \(\epsilon\to0\), the unbiased condition at \(\theta\) and \(\theta+\epsilon\) is equivalent to the local unbiased condition.

If we consider the inequality Eq.~\eqref{eq:lower_bounds} with the parallel amplitudes for two infinitesimally neighboring parameter points, it follows that
\begin{equation} 
	\mathcal E_\theta \geq \qty[ 4 \lim_{\epsilon\to0} 
		\frac{ d_\mathrm{B}(\rho_\theta,\rho_{\theta+\epsilon})^2}{\epsilon^2}
	]^{-1}
	= \frac1{\qfiThird_\theta}.
\end{equation}
Therefore, we obtain the QCRB without resorting to the SLD operator.

The purification method is often used for approaching the QCRB for noisy quantum-enhanced metrology, e.g., see Refs.~\cite{Fujiwara2008,Escher2011,Yuan2017}.
Here, the inequalities derived in Proposition~\ref{prop:purification} can take the range of the unbiased conditions from the local infinitesimal neighborhood to any two reference points.
This may be useful in further study on noisy-enhanced metrology.

\section{Generalization of the Yang-Chiribella-Hayashi inequality to biased estimators}
\label{app:YCH}
We here show that the original proof~\cite{Yang2019a} of the Yang-Chiribella-Hayashi inequality Eq.~\eqref{eq:YCH} can be slightly changed to include biased estimators.
For simplicity, we assume that the measurement outcome \(x\) is discretely valued.
The derivation can be easily generalized to the continuous valued observables.

First, it can be shown from the definitions of the variance and the mean of the estimator that
\begin{align}
	&\quad \mathcal V_\theta[\hat\theta] + \mathcal V_{\theta+\epsilon}[\hat\theta] 
		+ (\mathbb{E}_{\theta+\epsilon}[\hat\theta] - \mathbb{E}_\theta[\hat\theta])^2 \nonumber \\
	&= \sum_x \qty(\hat{\theta} - \mathbb{E}_\theta[\hat{\theta}])^2 \qty[p_{\theta}(x)+p_{\theta+\epsilon}(x)]. \label{eq:YCH1}
\end{align}
The following steps are similar to that in Ref.~\cite{Yang2019a}.
It follows from \(p_\theta(x) + p_{\theta+\epsilon}(x) \geq \qty[ \sqrt{p_\theta(x)} + \sqrt{p_{\theta+\epsilon}(x)} ]^2 / 2\) that
\begin{align}
	& \sum_x \qty(\hat\theta - \mathbb E_\theta[\hat\theta])^2 \qty[ p_\theta(x) + p_{\theta+\epsilon}(x) ] \nonumber\\
	\geq &\frac12 \sum_x \qty(\hat\theta - \mathbb E_\theta[\hat\theta])^2 
		\qty[ \sqrt{p_\theta(x)} + \sqrt{p_{\theta+\epsilon}(x)} ]^2.
	\label{eq:YCH2}
\end{align}
With the Bhattacharyya coefficient \(B(p_\theta,p_{\theta+\epsilon})	:= \sum_x \sqrt{p_\theta(x) p_{\theta+\epsilon}(x)}\), we have
\begin{equation}
	2[1-B(p_\theta,p_{\theta+\epsilon})] = \sum_x \qty[\sqrt{p_{\theta}(x)} - \sqrt{p_{\theta+\epsilon}(x)}]^2.
\end{equation}
It then follows from the Cauchy-Schwarz inequality that
\begin{align}
	&\quad \mbox{r.h.s. of Eq.~\eqref{eq:YCH2}} \times 2 [1 - B(p_\theta,p_{\theta+\epsilon})] \nonumber \\
	&\geq \frac12 \qty{ 
		\sum_x \qty(\hat{\theta}-\mathbb{E}_{\theta}[\hat{\theta}]) 
		\qty[p_{\theta}(x) - p_{\theta+\epsilon}(x)]
	}^2 \nonumber \\
	&= \frac12 \qty(\mathbb E_{\theta + \epsilon}[\hat\theta] - \mathbb E_{\theta}[\hat\theta])^2. 
	\label{eq:YCH3}
\end{align}
Combining Eqs.~\eqref{eq:YCH1}, \eqref{eq:YCH2}, \eqref{eq:YCH3}, and \(B(p_\theta,p_{\theta+\epsilon}) \geq \norm{\sqrt{\rho_\theta} \sqrt{\rho_{\theta+\epsilon}}}_1\) (see Ref.~\cite{Fuchs1999}), it can be shown that
\begin{align}
	&\quad \mathcal V_\theta[\hat\theta] + \mathcal V_{\theta+\epsilon}[\hat\theta]  
	+ (\mathbb{E}_{\theta+\epsilon}[\hat\theta] - \mathbb{E}_\theta[\hat\theta])^2 \nonumber \\
	&\geq \frac{\qty(\mathbb E_{\theta + \epsilon}[\hat\theta] - \mathbb E_{\theta}[\hat\theta])^2}
	{2 d_{\mathrm{B}}(\rho_\theta,\rho_{\theta+\epsilon})^2}.
\end{align}
For unbiased estimators, the above inequality becomes the Yang-Chiribella-Hayashi inequality Eq.~\eqref{eq:YCH}.
Taking the limit \(\epsilon\to0\) in the above inequality, we get the biased version of \(\qfiThird\)-QCRB:
\begin{equation}
	\mathcal V_\theta 
	\geq \qty( \dv{\theta} \mathbb E_\theta[\hat\theta] )^2 / \qfiThird_\theta.
\end{equation}

\bibliography{../../../Workplace/wiki}

\end{document}